\documentclass[10pt,a4paper,twocolumn,aps,prd,notitlepage,nofootinbib,showpacs]{revtex4-1}
\usepackage[latin9]{inputenc}
\usepackage{float}
\usepackage{amsmath}
\usepackage{graphicx}

\makeatletter

\pdfpageheight\paperheight
\pdfpagewidth\paperwidth

\usepackage{color}

\allowdisplaybreaks

\newcommand{\be}{\begin{equation}}
\newcommand{\ee}{\end{equation}}

\makeatother

\begin{document}

\title{General-relativistic rotation: Self-gravitating fluid tori in motion
around black holes}

\author{Janusz Karkowski}

\author{Wojciech Kulczycki}

\author{Patryk Mach}

\author{Edward Malec}

\author{Andrzej Odrzywo\l ek}

\author{Micha\l ~ Pir\'og}

\affiliation{Instytut Fizyki im.~Mariana Smoluchowskiego, Uniwersytet Jagiello\'{n}ski,
{\L }ojasiewicza 11, 30-348 Krak\'ow, Poland}
\begin{abstract}
We obtain from the first principles a general-relativistic Keplerian
rotation law for self-gravitating disks around spinning black holes.
This is an extension of a former rotation law that was designed mainly
for toroids around spinless black holes. We integrate numerically
axial stationary Einstein equations with self-gravitating disks around
spinless or spinning black holes; that includes the first ever integration
of the Keplerian selfgravitating tori. This construction can be used
for the description of tight black hole-torus systems produced during
coalescences of two neutron stars or modelling of compact active galactic
nuclei. 
\end{abstract}

\maketitle

\section{Introduction}

Keplerian rotation is common in rotating astrophysical systems. Quasistationary
fluid tori rotating around black holes may emerge in binary neutron
stars or black hole-neutron star mergers (\cite{BR2017} and references
therein). There is a recent detection of gravitational waves coming
from such a binary merger \cite{GW17082017}. Tori in these systems
seem to be relatively massive and yet rotate with the Keplerian velocity
\cite{NSkeplerian,KR,ECGKK,SFHKKST}. We proposed in \cite{MM} a
general-relativistic (GR in what follows) differential rotation law
capable of describing a stationary system consisting of a self-gravitating
disk circulating around a spinless or spinning black hole (BH hereafter).
Its interesting feature (absent in other proposals \cite{Butterworth_Ipser,Bardeen_1970,GYE,UTGHSTY,Uryu,Ansorg_Petroff})
is that the nonrelativistic limit exactly yields the Newtonian angular
velocity $\Omega=w/d^{\lambda}$, where $w$ is a constant. Here $0\le\lambda\le2$,
and $d$ is the Euclidean distance from the rotation axis. This limit
includes the Keplerian rotation law with $\lambda=3/2$. This rotation
law has been recently investigated within the first post-Newtonian
(1PN) approximation, for a spinless BH \cite{KMMPX}.

The main goals of this paper are: (i) to show a new rotation law that
can be more adequate for Keplerian systems consisting of a spinning
BH and a disk; (ii) describe numerically a polytropic disk, using
both versions of rotation laws\textemdash that of \cite{MM} and the
new one\textemdash in the full Einstein theory within the puncture
framework as formulated by Shibata \cite{MSH}.

We shall give a concise report on the formalism and obtained results;
details are given in a longer accompanying paper \cite{KKMMOP}. In
numerical calculations we investigated several values of the spin
parameter in the interval $a\in[-0.9,0.9]$ and mainly light disks,
that include cases shown to be of interest in recent simulations of
coalescences of neutron stars \cite{BR2017,Kastaun2013}.

\section{Equations}

We choose a stationary metric 
\begin{align}
ds^{2} & =-\alpha^{2}dt^{2}+r^{2}\sin^{2}\theta\psi^{4}\left(d\phi-\beta dt\right)^{2}\nonumber \\
 & \qquad+\psi^{4}e^{2q}\left(dr^{2}+r^{2}d\theta^{2}\right).
\end{align}
We put $G=c=1$. Here $t$ is the time coordinate, and $r$, $\theta$,
$\phi$ are spherical coordinates. We assume axial symmetry and employ
the stress-momentum tensor $T^{\alpha\beta}=\rho hu^{\alpha}u^{\beta}+pg^{\alpha\beta},$
where $\rho$ is the baryonic rest-mass density, $h$ is the specific
enthalpy, and $p$ is the pressure. We assume that $p(\rho)=K\rho^{\frac{4}{3}}$,
but our analysis can be done for any barotropic equation of state.
The 4-velocity $(u^{\alpha})=(u^{t},0,0,u^{\phi})$ is normalized,
$u_{\beta}u^{\beta}=-1$. The angular velocity $\Omega=u^{\phi}/u^{t}$.
We use Einstein-hydrodynamic equations as formulated by Shibata (see
Sec.\ II in \cite{MSH} or Secs.\ II and III in \cite{KKMMOP}).
It is well known that GR Euler equations are solvable under an integrability
condition \cite{Butterworth_Ipser,Bardeen_1970}. In accordance with
that, we assume that the angular momentum per unit inertial mass $\rho h$
\cite{FM}, $j\equiv u_{\phi}u^{t}$ depends only on the angular velocity
$\Omega$, $j\equiv j(\Omega)$.

The metric functions satisfy equations (44)\textendash (48) in \cite{MSH},
with Shibata's boundary conditions on the horizon of the central BH,
at the symmetry axis and at spatial infinity. We used the Kerr metric\textemdash the
only available analytic solution\textemdash for testing the correctness
of numerical codes. That led to a somewhat different numerical implementation
of some of the boundary conditions of \cite{MSH}, in order to get
the Kerr solution with the best accuracy in the absence of the torus
(see \cite{KKMMOP} for details).

The BH is surrounded by a minimal two-surface $S_{\mathrm{BH}}$ (on
a fixed hypersurface of constant time). Its area $A_{\mathrm{H}}$
defines the irreducible mass $M_{\mathrm{irr}}=\sqrt{\frac{A_{\mathrm{H}}}{16\pi}}$
and its angular momentum $J_{\mathrm{H}}$ follows from the Komar
expression $J_{\mathrm{H}}=\frac{1}{4}\int_{0}^{\pi/2}\frac{r^{4}\psi^{6}}{\alpha}\partial_{r}\beta\sin^{3}\theta d\theta$.
The angular momentum is prescribed rigidly on the event horizon $S_{\mathrm{BH}}$\textemdash it
is given by data taken from the Kerr solution (with two parameters:
mass $m$ and the spin parameter $a=J_{\mathrm{H}}/m$), and it is
independent of the content of mass in the torus. The mass of the BH
is then defined as $M_{\mathrm{BH}}=M_{\mathrm{irr}}\sqrt{1+\frac{J_{\mathrm{H}}^{2}}{4M_{\mathrm{irr}}^{4}}}$.
Another possible choice of the BH mass is in terms of the circumferential
radius $r_{\mathrm{C}}$ of $S_{\mathrm{BH}}$ at the symmetry plane
$\theta=\pi/2$: $M_{\mathrm{C}}=r_{\mathrm{C}}/2$. We observed that
in numerical calculations $M_{\mathrm{C}}$ and $M_{\mathrm{BH}}$
differ by significantly less than 1\% \cite{KKMMOP}; this is consistent
with findings of Shibata \cite{MSH}. The asymptotic mass $M_{\mathrm{ADM}}$
is defined as in \cite{MSH}. The mass of tori is defined by $M_{\mathrm{T}}\equiv M_{\mathrm{ADM}}-M_{\mathrm{BH}}$.

In \cite{MM} we have got the rotation law for the motion around central
bodies 
\begin{equation}
j(\Omega)\equiv\frac{\tilde{w}^{1-\delta}\Omega^{\delta}}{1-\kappa\tilde{w}^{1-\delta}\Omega^{1+\delta}+\Psi};\label{primordial}
\end{equation}
here $\Psi$ is of the order of the binding energy per unit baryonic
mass, and $\kappa=\frac{1-3\delta}{1+\delta}$. This formula followed
from an ``educated guesswork'', with three basic elements: (i) the
rotation law should have the right (monomial) form in the Newtonian
limit; (ii) the 1PN correction to the Bernoulli equation should have
the right form; (iii) the first post-Newtonian (1PN) correction to
the angular velocity should be uniquely defined. The 1PN analysis
of the Bernoulli equation yields that (\ref{primordial}) is admissible
for any $\kappa$. The fact that $\kappa$ can be arbitrary, is unfortunate,
because its values influence the angular velocity $\Omega$\textemdash a
quantity that can be observable. There exists, however, a physical
constraint, that follows from the fact that a massless disk of dust
is just a collection of test particles in a circular motion. Therefore
such a disk in the Schwarzschild geometry must exactly satisfy the
Einstein-Bernoulli equations, with the angular velocity $\Omega=\sqrt{M_{\mathrm{BH}}}/r_{\mathrm{C}}^{3/2}$.
That implies $\kappa=\frac{1-3\delta}{1+\delta}$, and the relation
between $\kappa$ and $\delta$ becomes unique. The full reasoning
is described in \cite{MM}. Let us remark that spins of BH's do not
appear in the 1PN approximation, and therefore (\ref{primordial})
might be expected to apply only to systems with spin-less BH's. Numerical
analysis suggests that it is valid also for spinning BH's (see below).
Simple rescaling $w^{1-\delta}=\frac{\tilde{w}^{1-\delta}}{1+\Psi}$
transforms (\ref{primordial}) into 
\begin{equation}
j(\Omega)\equiv\frac{w^{1-\delta}\Omega^{\delta}}{1-\kappa w^{1-\delta}\Omega^{1+\delta}},\label{momentum}
\end{equation}
where $w$, $\delta$ and $\kappa=\frac{1-3\delta}{1+\delta}$ are
parameters \cite{Krolik}. Note that only the value of $\delta\in(-\infty,0)$
is arbitrary, and that $w$ is obtained as a part of the solution.
The cases $\delta=-1$, $\delta=-1/3$ and $\delta=0$ correspond\textemdash in
the Newtonian limit\textemdash to the constant linear velocity, the
Keplerian rotation, and the constant angular momentum density, respectively.
Let us stress, that in this construction the central body can be a
spinless BH, but equally well there can be no central body. We believe
also, that this differential rotation can be adapted to rotating stars.
The rotation law (\ref{momentum}) can be understood as the general-relativistic
version of the Poincar\'{e}-Wavre condition for fluids undergoing monomial
rotation, $\Omega=w/d^{\lambda}$, in Newtonian gravity.

We performed dozens of numerical runs with different pairs $(\delta,\kappa=\frac{1-3\delta}{1+\delta})$
(where $\delta\in[-0.99,0]$) and, remarkably, there always existed
solutions. This is reported below.

The rotation curves $\Omega\left(r,\cos\theta\right)$ ought to be
recovered from $j(\Omega)=u_{\phi}u^{t}$: \be \label{rotation_law}
\frac{w^{1-\delta } \Omega^\delta }{1- \kappa w^{1- \delta
}\Omega^{1+\delta } } = \frac{V^2}{\left( \Omega +\beta
\right) \left( 1- V^2 \right) }, \ee where the square of the linear
velocity reads $V^{2}=r^{2}\sin^{2}\theta\left(\Omega+\beta\right)^{2}\frac{\psi^{4}}{\alpha^{2}}.$
The Euler equations reduce to a GR integro-algebraic Bernoulli equation,
that embodies the hydrodynamic information carried by the continuity
equations $\nabla_{\mu}T^{\mu\nu}=0$ and the baryonic mass conservation
$\nabla_{\mu}\left(\rho u^{\mu}\right)=0$. It becomes \cite{MM}

\begin{eqnarray}
h\alpha\sqrt{1-V^{2}}\left(1-\kappa w^{1-\delta}\Omega^{1+\delta}\right)^{-\frac{1}{\left(1+\delta\right)\kappa}}=C.\label{algebraic_Bernoulli}
\end{eqnarray}

\section{Tori in the Keplerian rotation around spinning black holes}

Below we obtain analytically a generalization of the rotation curve
(\ref{momentum}) that describes tori in motion around spinning BH's.
It is known that test particles can rotate along circular orbits $r=\mathrm{const}$
in the Kerr geometry. That means that there exists a testlike disk
made of dust, that moves circularly and lies on the plane $\theta=\pi/2$.
The radial variable $r$ of each disk particle can be expressed in
terms of its angular velocity $\Omega(r)=\frac{8r^{3/2}}{\left((2r+1)^{2}-a^{2}\right)^{3/2}+8ar^{3/2}}$.
Metric functions $\psi,\beta,\alpha,q$ of the Kerr geometry depend
only on the variable $r$, on the disk's surface, hence they can be
expressed entirely in terms of $\Omega$. Thus the angular momentum
density 
\begin{eqnarray}
j_{\mathrm{K}}=u_{\phi}u^{t}=\frac{r^{2}\left(\Omega+\beta\right)\frac{\psi^{4}}{\alpha^{2}}}{1-\left(\Omega+\beta\right)^{2}\frac{r^{2}\psi^{4}}{\alpha^{2}}}\label{jk}
\end{eqnarray}
becomes a function $j_{\mathrm{K}}(\Omega)$ of the angular velocity
$\Omega$. Let us now define a function $F(\Omega)$ as a solution
of 
\begin{equation}
\frac{d}{d\Omega}\ln(1-F(\Omega))=-2j_{\mathrm{K}}(\Omega);\label{rotation_spin}
\end{equation}
from that we get after lengthy calculations \cite{KKMMOP} 
\begin{equation}
F(\Omega)=a^{2}\Omega^{2}+3m^{\frac{2}{3}}\Omega^{\frac{2}{3}}(1-a\Omega)^{\frac{4}{3}}.\label{F}
\end{equation}
Here $m$ is the mass of the Kerr BH and the formula describes circular
motions of massless disks of dust. The parameter $a$ in (\ref{F})
is just the spin parameter of the Kerr BH.

After finding (\ref{rotation_spin}), we discovered a much simpler
derivation in Boyer-Lindquist coordinates (the radial coordinate $r_{\mathrm{BL}}$
is different from that used before). The angular velocity reads $\Omega=\frac{\sqrt{m}}{\pm r_{\mathrm{BL}}^{3/2}+a\sqrt{m}}$
(formula (2.16) in \cite{Bardeen1972}); that gives $r_{\mathrm{BL}}=\frac{m^{1/3}(1-a\Omega)^{2/3}}{\Omega^{2/3}}$.
Replacing $r_{\mathrm{BL}}$ in $j=u_{\phi}u^{t}=-\frac{g_{0\phi}+g_{\phi\phi}\Omega}{g_{00}+2g_{0\phi}\Omega+g_{\phi\phi}\Omega^{2}}$,
one arrives at (\ref{rotation_spin}) and (\ref{F}).

The application of this approach to the circular motion of the massless
disk of dust in Schwarzschild geometry yields $r_{\mathrm{C}}=w^{2/3}\Omega^{-2/3}$
and $F=3w^{4/3}\Omega^{2/3}$; this agrees with (\ref{momentum})
for the Keplerian motion.

The rotation curves $\Omega\left(r,\cos\theta\right)$ ought to be
found from \be \label{rotation_lawS} -\frac{3}{2\kappa} \frac{d}{d\Omega
} \ln \left(1- \frac{\kappa }{3} F(\Omega) \right) = \frac{V^2}{\left(
\Omega +\beta \right) \left( 1- V^2 \right) }. \ee It is clear
from the construction that test disks of dust in the Kerr geometry
(with $\kappa=3$) satisfy Eq.\ (\ref{rotation_lawS}), Einstein
equations and the Bernoulli equation 
\begin{equation}
h\alpha\sqrt{1-V^{2}}\left(1-\frac{\kappa}{3}F(\Omega)\right)^{-1/2}=C.\label{Bernoulli2}
\end{equation}

We conjecture that 
\begin{equation}
j(\Omega)\equiv-\frac{3}{2\kappa}\frac{d}{d\Omega}\ln \left(1-\frac{\kappa}{3}(a^{2}\Omega^{2}+3w^{\frac{4}{3}}\Omega^{\frac{2}{3}}(1-a\Omega)^{\frac{4}{3}}) \right) \label{rotation_spin1}
\end{equation}
is the rotation law for a heavy torus rotating around a black hole
with the spin parameter equal to the parameter $a$ that appears in
(\ref{rotation_spin1}).

$\kappa$ might be regarded as a free parameter in the rotation law
(\ref{rotation_spin1}), but its set of allowed values centers closely
around $3$. Moreover, the choice $\kappa=3$ always yielded solutions
in our numerical calculations. The parameter $w$ in turn is the only
unknown and it can be obtained in the process of solving the problem.

Our numerical method follows closely the scheme proposed in \cite{MSH},
with only few modifications \cite{KKMMOP}. The most important one
is due to a different rotation law\textemdash the replacement of one
rotation curve by another in the numerical programme can be quite
difficult and might require considerable work. Equation (11) of \cite{MSH}
is replaced with our Eq.\ (\ref{algebraic_Bernoulli}), which is
equivalent to Eq.\ (A11) of \cite{MSH}, and which suits better the
convention with $j=u_{\phi}u^{t}$. Our implementation is described,
with a detailed account of the boundary conditions, in \cite{KKMMOP}.

The solution is specified by setting 7 main parameters: the mass $m$
($m=1$ in forthcoming examples) and spin $a$ defined as in \cite{MSH}\textemdash they
describe the BH; the inner and outer coordinate radii of the disk
at the equatorial plane: $r_{1}$ and $r_{2}$, respectively; the
maximum of the rest-mass density within the disk $\rho_{0}$; parameters
$\delta$ and $\kappa$ in (\ref{momentum}). We check the virial
condition of \cite{MSH}\textemdash that the Komar and asymptotic
masses are equal. We performed several tests of numerical codes \cite{KKMMOP}.

\section{Numerical results: tori around spinless black holes, rotation curve
(\ref{momentum})}

\begin{figure}[h]
\includegraphics[width=1\columnwidth]{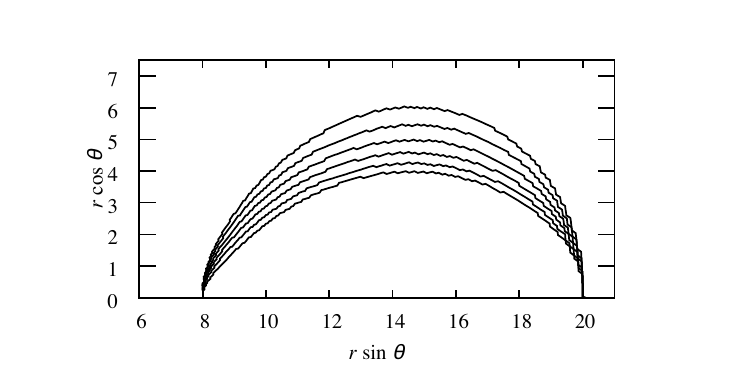}\caption{ Toroidal profiles for rotation around spin-less BH's, rotation law
(\ref{momentum}) with $\delta=-0.99,-0.8,-0.6,-0.4,-0.2,0$; toroids
flatten with the decrease of $\delta$. Masses: $M_{\mathrm{BH}}\in(1.076,1.079)$;
$M_{\mathrm{ADM}}=2.0$. Here $r_{1}=8$ and $r_{2}=20$. }
\label{Fig1} 
\end{figure}

We shall describe 6 solutions, with the following pairs of parameters,
$(\delta,\kappa=\frac{1-3\delta}{1+\delta})$: $[(-0.99,397);(-0.8,17);(-0.6,7);(-0.4,11/3);(-0.2,2);$
$(0,1)]$. It is remarkable, that they exist exactly at sets of numbers
($\delta,\kappa(\delta)$) predicted by \cite{MM}\textemdash Eq.\ (\ref{momentum})
does not have any free parameters in the case of spinless BH's. Profiles
of the obtained tori are shown in Fig.\ \ref{Fig1}. They flatten
with the decrease of the parameter $\delta$, maximal values of their
mass density increase, by a factor of 2, (from $0.000454$ at $\delta=0$
to $0.000879$ at $\delta=-0.99$) while their maximal specific enthalpy
decreases rather moderately (from $1.0348$ at $\delta=0$ to $1.0273$
at $\delta=-0.99$).

Table \ref{table1} shows a sample of results concerning the Keplerian
rotation {[}Eq.\ (\ref{momentum}) with $\delta=-1/3${]} around
spinless BH's.

With the increase of mass, while keeping the inner boundary close
to the innermost stable circular orbit, the interval $(\kappa_{\mathrm{min}},\kappa_{\mathrm{max}})$
for which solutions were found, shifts downwards: both bounds for
$\kappa$ go down. The choice of $\kappa=3$ always gives solutions.

%[H]

Empirical experience, gained through numerical investigations, shows
that the rotation law (\ref{rotation_spin1}) is valid beyond the
range of applicability that is suggested by the preceding derivation.
One can consider fluid tori in the Keplerian motion around BH's with
the spin parameter $a$, and the rotation curve (\ref{rotation_spin1}):
\begin{equation}
j(\Omega)\equiv-\frac{1}{2}\frac{d}{d\Omega}\ln(1-(b^{2}\Omega^{2}+3w^{\frac{4}{3}}\Omega^{\frac{2}{3}}(1-b\Omega)^{\frac{4}{3}})\label{jb}
\end{equation}
with $b$ that might be different from $a$. All pairs $(a,b)\in(-1,1)\times(-1,1)$
appear to be admissible (but see a discussion below). Setting $b=0$
in the pair $(a,b)$ ($a\ne0$) converts (\ref{rotation_spin1}) into
(\ref{momentum}); but this corresponds to the Keplerian motion of
tori around spinning (if $a\ne0$) or spinless black holes ($a=0$).
Thus the Keplerian version of the rotation law (\ref{momentum}) is
valid also for spinning BH's.

\begin{table}[h]
\caption{\label{table1} The first 2 columns give the innermost and outermost
radii of tori, the third describes maximal values of the baryonic
mass density, the fourth and fifth give $M_{\mathrm{T}}$ and $M_{\mathrm{BH}}$
respectively (with $\kappa=3$), and the final two columns show the
minimal and maximal values of $\kappa$ for which solutions were found. }
\begin{ruledtabular}
\begin{tabular}{ccccccc}
$r_{1}$  & $r_{2}$  & $\rho_{0}$  & $M_{\textrm{T}}$  & $M_{\textrm{BH}}$  & $\kappa_{\mathrm{min}}$  & $\kappa_{\mathrm{max}}$ \\ \hline
$5$  & $20$  & $0.2\times10^{-4}$  & $0.005$  & $1.000$  & $3.0$  & $4.9$ \\
$5$  & $20$  & $1.0\times10^{-4}$  & $0.078$  & $1.008$  & $2.9$  & $4.8$ \\
$5$  & $20$  & $5.0\times10^{-4}$  & $1.1$  & $1.120$  & $1.5$  & $4.0$ \\
$5$  & $10$  & $0.2\times10^{-3}$  & $0.007$  & $1.001$  & $3.0$  & $5.2$ \\
$5$  & $10$  & $1.0\times10^{-3}$  & $0.11$  & $1.017$  & $2.7$  & $5.0$ \\
$5$  & $10$  & $4.0\times10^{-3}$  & $1.0$  & $1.176$  & $0.8$  & $3.9$ \\
\end{tabular}
\end{ruledtabular}
\end{table}

\section{Numerical results: tori around spinning BH's, rotation curve (\ref{momentum})
with $\boldsymbol{\delta=-0.5}$}

\begin{figure}[h]
\includegraphics[width=1\columnwidth]{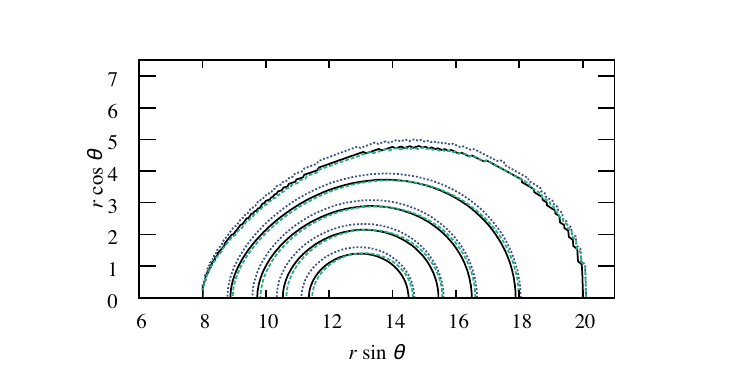}\caption{ Toroidal profiles for rotation around spinning BH's, rotation law
(\ref{momentum}) with $\delta=-0.5$ and spins $a=-0.5$ (broken
green line, $M_{\mathrm{T}}=0.948$), $a=0.0$ (solid black line,
$M_{\mathrm{T}}=0.922$), $a=0.5$ (dotted blue line, $M_{\mathrm{T}}=0.953$).
The asymptotic mass $M_{\mathrm{ADM}}=2.0$. The innermost and outermost
coordinate radii are $8$ and $20$, respectively. The internal isolines
correspond to $h=1.02398,1.01799,1.01199,1.006$, with the outermost
isoline representing $h=1+10^{-15}$.}
\label{Fig2} 
\end{figure}

We analyzed toroids circulating around BH's with the spin parameters
$a=-0.5,0.0,0.5$. Resulting profiles are shown in Fig.\ \ref{Fig2};
their shape depends rather weakly on spins. The enthalpy and mass
density isolines are relatively insensitive to spins of BH's, similarly
as their maximal values, which change by at most 5\%.

%[H]

\begin{table}[H]
\caption{\label{table3} Minimal masses of Keplerian tori for $\kappa=3$ and
$b=0$. Here $r_{1}=8$, $r_{2}=20$ and $M_{\mathrm{BH}}\in(1.001,1003)$. }

\begin{ruledtabular}
\begin{tabular}{cccccccccc}
$a$  & $-0.9$  & $-0.8$  & $-0.7$  & $-0.6$  & $-0.5$  & $-0.4$  & $-0.3$  & $-0.2$  & $-0.1$ \\
$M_{\mathrm{T}}$  & 0.143  & 0.127  & 0.111  & 0.086  & 0.075  & 0.057  & 0.045  & 0.034  & 0.014 \\
\end{tabular}
\end{ruledtabular}

\end{table}

\begin{table}[H]
\caption{\label{table4} Minimal masses of Keplerian tori for $\kappa=3$ and
$a=-0.5$. Here $r_{1}=5$, $r_{2}=20$. }

\begin{ruledtabular}
\begin{tabular}{ccccc}
$b$  & $-0.3$  & $-0.4$  & $-0.45$  & $-0.5$ \\
$M_{\mathrm{T}}$  & $6.23\times10^{-2}$  & $3\times10^{-2}$  & $1.52\times10^{-2}$  & 0 \\
\end{tabular}
\end{ruledtabular}

\end{table}

\section{The mass gap conjecture}

Consider a BH with the spin parameter $a$ and a torus evolved according
to (\ref{jb}). Assume $\kappa=3$ and $\Omega>0$. Then there exists
a mass gap in the spectrum of toroidal masses, if $b-a\ge0$.

We show here two classes of numerical data that support this conjecture.

\begin{enumerate}
\item [(i)] The spin parameter $a$ is negative, $a\in[-0.1,-0.9]$, while
the rotation parameter vanishes, $b=0$. Then it appears that tori
may exist only if their mass is larger than a particular mass threshold.
This mass gap goes to zero with vanishing $a$ (see Table \ref{table3}).
\item [(ii)] $a=-0.5$ and $b=-0.5,-0.45,-0.4,-0.3$. Toroids exist only if
their mass is larger than a certain minimal value, which goes to zero
with vanishing $b-a$ (Table \ref{table4}).
\end{enumerate}

The existence of this mass gap is of interest from the mathematical
point of view, since it means that the spacetime geometry produced
by counter-rotating disks with the above specified rotation law does
not tend to the Kerr geometry; solutions cease to exist when masses
of toroids become too small. We have found that this mass threshold
disappears for other values of $\kappa$ \cite{KKMMOP}.

\section{Numerics: tori around spinning BH's, rotation curve (\ref{rotation_spin1})}

\begin{figure}[b]
\includegraphics[width=1\columnwidth]{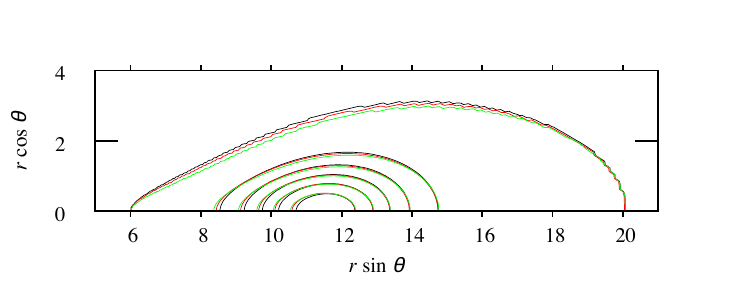}\caption{ Toroidal mass density profiles ($\rho=10^{-12}$) and isolines $\rho=0.00002,0.00004,0.00006,0.00008,0.0001$
for the rotation law (\ref{rotation_spin1}), $a=b=$: $-0.9$ (solid
line); $0$ (dotted line); $0.9$ (dashed line). Masses: $M_{\mathrm{BH}}\in[1.001,1.009]$;
$M_{\mathrm{ADM}}\in[1.101,1.110]$; $M_{\mathrm{T}}\approx0.10$.
Here $r_{1}=6$ and $r_{2}=20$. }
\label{Fig3} 
\end{figure}

Figure \ref{Fig3} demonstrates that shapes of tori, their enthalpy
and mass density, depend weakly on the value of the spin parameter
$a$. The mass spectrum of tori starts from zero, as expected, since
now $b-a=0$.

%[H]

Both formulae (\ref{momentum}) and (\ref{rotation_spin1}) can be
understood as a prescription for the spin-orbit interaction between
BH's and fluid tori around them. The new rotation law of this letter
can be more suitable than the former version \cite{MM} for the evolution
of Keplerian disks around spinning BH's, since it gives the gapless
mass spectrum of tori and does not contain any free parameters.

We have shown that the former law \cite{MM} can be numerically implemented
within the puncture framework \cite{MSH}. It is capable to describe
compact BH-torus systems that can be created in the merger of compact
binaries consisting of pairs of neutron stars \cite{NSkeplerian,SFHKKST,Pan_Ton_Rez,Lovelace}
and that might exist also in some galactic centers \cite{Bardeen-Petterson,Moran}.
Our results agree with the earlier post-Newtonian analysis (see \cite{KMMPX,KMM,JMMP});
the relevant analysis will be published elsewhere. Self-gravitating
fluid bodies can now be investigated in the regime of strong gravity
for GR versions of the Keplerian rotation, including stationary disks
in tight accretion systems with central (spinless or spinning) BH's
or rotating stars (with some adjustments of the rotation law in the
vicinity of the symmetry axis). Our construction can be used in order
to specify initial data for disk-BH configurations with viscosity;
their evolution would model formation of ejecta in post-merger remains
of two coalescing neutron stars.
\begin{acknowledgments}
This research was carried out with the supercomputer ``Deszno''
purchased thanks to the financial support of the European Regional
Development Fund in the framework of the Polish Innovation Economy
Operational Program (contract no.\ POIG.\ 02.01.00-12-023/08). M.P.
acknowledges partial support from the Grant No. K/DSC/004356. P.M.
acknowledges the financial support of the Narodowe Centrum Nauki Grant
No. DEC-2012/06/A/ST2/00397. 
\end{acknowledgments}

\end{document}